\documentclass[useAMS,usenatbib,usegraphicx,letterpaper]{mn2e}
\usepackage{times,amssymb,hyperref,aas_macros}

\usepackage[totalwidth=480pt,totalheight=680pt,layoutvoffset=0.5cm]{geometry}

\begin{document}

\title[ALMA observations of HR~4796A]{ALMA observations of the narrow HR~4796A debris ring}

\author[Grant M. Kennedy et al.]{Grant M. Kennedy\thanks{Email:
    \href{mailto:g.kennedy@warwick.ac.uk}{g.kennedy@warwick.ac.uk}}$^1$,
    Sebastian Marino$^2$, Luca Matr\`a$^3$, Olja Pani\'c$^4$, \newauthor
    David Wilner $^3$, Mark C. Wyatt$^2$, Ben Yelverton$^2$ \\
  $^1$ Department of Physics, University of Warwick, Gibbet Hill Road, 
Coventry, CV4 7AL, UK \\
 $^2$ Institute of Astronomy, University of Cambridge, Madingley Road,
Cambridge CB3   0HA, UK \\
  $^3$ Harvard-Smithsonian Center for Astrophysics, 60 Garden Street,
Cambridge, MA   02138, USA \\
  $^4$ School of Physics and Astronomy, University of Leeds, Leeds LS2 9JT, UK \\
 }

\maketitle

\begin{abstract}
  The young A0V star HR~4796A is host to a bright and narrow ring of
  dust, thought to originate in collisions between planetesimals within
  a belt analogous to the Solar System's Edgeworth-Kuiper belt. Here we
  present high spatial resolution 880$\mu$m continuum images from the
  Atacama Large Millimeter Array. The 80au radius dust ring is resolved
  radially with a characteristic width of 10au, consistent with the
  narrow profile seen in scattered light. Our modelling consistently
  finds that the disk is also vertically resolved with a similar
  extent. However, this extent is less than the beam size, and a disk
  that is dynamically very cold (i.e. vertically thin) provides a better
  theoretical explanation for the narrow scattered light profile, so we
  remain cautious about this conclusion. We do not detect $^{12}$CO
  J=3-2 emission, concluding that unless the disk is dynamically cold
  the CO+CO$_2$ ice content of the planetesimals is of order a few
  percent or less. We consider the range of semi-major axes and masses
  of an interior planet supposed to cause the ring's eccentricity,
  finding that such a planet should be more massive than Neptune and
  orbit beyond 40au. Independent of our ALMA observations, we note a
  conflict between mid-IR pericenter-glow and scattered light imaging
  interpretations, concluding that models where the spatial dust density
  and grain size vary around the ring should be explored.
\end{abstract}


\section{Introduction}\label{s:intro}

The belts of asteroids and comets that orbit the Sun and other stars
have long been recognised as tracers of system-wide dynamics, and thus
used as a means to discover perturbations from unseen planets
\citep[e.g.][]{1997MNRAS.292..896M,2005Natur.435.1067K}. Indeed, much of
the history of how these planetesimal belts --- the so-called `debris
disks' --- have been studied is the application of Solar System
dynamics to other stars.

These ideas can be broadly split into the short and long-term effects of
planets on the appearance of a disk. The former is usually related to
resonances and produces small-scale `clumpy` dust structure
\citep[e.g.][]{1999AJ....118..580L,2003ApJ...598.1321W}. The latter can
be thought of as the perturbations induced if a planet is smeared out
around its orbit, and produces large-scale structures such as eccentric
rings and warps \citep[e.g.][]{1997MNRAS.292..896M,1999ApJ...527..918W}.
Structures consistent with the long-term (`secular') perturbations
have been robustly detected and quantified in a number of systems \citep
[e.g.][]{2005Natur.435.1067K,2006AJ....131.3109G,2011A&A...526A..34M},
but whether clumps have ever been detected in a debris disk is
debatable; for example the azimuthal structure reported in various
mm-wave images of $\epsilon$ Eridani's disk
\citep[e.g.][]{2005ApJ...619L.187G,2015A&A...576A..72L} has not been
detected in others with comparable or greater depth
\citep{2015ApJ...809...47M,2016MNRAS.462.2285C}. The best candidate for
clumpy disk structure is $\beta$ Pictoris, though the edge-on geometry
hinders deprojection of the disk to derive the spatial dust (and gas)
distribution \citep[e.g.][]{2014Sci...343.1490D}.

To successfully discern the spatial structure of these belts, and thus
search for evidence of planetary influence, requires images. While
debris disks are discovered by infrared (IR) flux densities that are in
excess of that expected from their host stars, our ability to infer even
basic radial disk structure from the disk spectrum is extremely poor.
While two sufficiently well separated belts can be distinguished from a
single narrow belt \citep{2014MNRAS.444.3164K}, whether these two belts
are really a single wide belt, and at what specific distance these belts
reside is almost always unknown \citep[see][for an early review on
inferring debris disk structure from spectra]{1993prpl.conf.1253B}.

The first debris disk to be imaged, around $\beta$ Pictoris
\citep{1984Sci...226.1421S}, showed a warp that was interpreted as
arising from a giant planet that is inclined to the disk by a few
degrees \citep{1995AAS...187.3205B,1997MNRAS.292..896M}, a planet that
has almost certainly now been detected \citep{2010Sci...329...57L}.
Subsequent images of other disks emerged 15 years later, at sub-mm
\citep[Fomalhaut and Vega,][]{1998Natur.392..788H} and mid-IR
wavelengths
\citep[HR~4796A,][]{1998ApJ...503L..79J,1998ApJ...503L..83K}. The disk
around HR~4796A was soon after shown to exhibit `pericenter glow'
\citep{1999ApJ...527..918W,2000ApJ...530..329T,2011A&A...526A..34M}.
With this phenomenon, mid-IR observations can detect a small but
coherent disk eccentricity, because the temperature increase for
particles at pericenter manifests as a large surface brightness
difference at wavelengths shorter than the peak flux (a similar effect
is an increased pericenter brightness in scattered light images). A
different manifestation of the same eccentricity is `apocenter glow',
where the apocenter of the same eccentric disk is brighter at
wavelengths longer than the peak, because the increase in dust density
outweighs the increase in temperature
\citep{2005A&A...440..937W,2016ApJ...832...81P}.  Both pericenter and
apocenter glow have now been detected for the disk around Fomalhaut
\citep{2005Natur.435.1067K,2012A&A...540A.125A,2017ApJ...842....8M}.

For $\beta$ Pic and HR~4796A, these observations are made at the 10 to
20 million-year age that places these systems just beyond the gas-rich
phase of planetesimal and planet construction, which always precedes the
ongoing destruction observed in debris disks. Systems at this age merit
study for myriad reasons; a few that are relevant here are:

\begin{enumerate}

\item Giant planets are brightest when they are youngest
  \citep[e.g.][]{1997ApJ...491..856B}, so detections are more likely and
  non-detections more constraining. Thus, direct imaging surveys focus
  on these stars.

\item Remnant gas from the protoplanetary phase may be present
  \citep{1995Natur.373..494Z,2011ApJ...740L...7M} and influence the disk
  structure in unexpected ways \citep{2013Natur.499..184L}. Quantifying
  the levels of gas (e.g. the dust/gas ratio) is important as it sets
  the context and the types of models used to interpret particular
  systems.

\item Debris disk mass, and thus brightness, decays with time
  \citep[e.g.][]{2003ApJ...598..636D,2005ApJ...620.1010R,2007ApJ...663..365W},
  so on average better images of disk structure can be obtained around
  younger stars (as long as the stars are not too distant).

\item A corollary of (iii) is that secondary gas released in
  planetesimal collisions, which depends on the disk's mass and yields
  compositional information, is more likely to be detected
  \citep{2015MNRAS.447.3936M}.

\item Secular perturbations have had less time to act on planetesimals,
  meaning that constraints on unseen perturbers, in concert with item
  (i), are stronger.

\end{enumerate}

Here we report the first Atacama Large Millimeter Array (ALMA)
observations of the narrow debris ring around HR~4796A (HD~109573,
HIP~61498, TWA~11A), an A0V star at 72.8~parsecs. The absolute
brightness of this disk, the 2\arcsec~diameter, and its location in the
southern hemisphere ($\delta=-40^\circ$) make this system perfectly
suited to the current generation of high-resolution optical and mm-wave
instruments. As a member of the $\sim$10 Myr-old TW~Hydrae association
\citep{1989ApJ...343L..61D,1997Sci...277...67K,1998ApJ...498..385S,1999ApJ...512L..63W,2015MNRAS.454..593B},
this system is young, so observations are well motivated for the reasons
listed above, and this system has been, and will continue to be, a
benchmark debris disk where theories can be tested in detail.

As a well studied system, there are a number of key results from the
prior study of this system. The aforementioned pericenter glow was the
first evidence that the disk is eccentric, and this has been
consistently confirmed with scattered light imaging
\citep{2009AJ....137...53S,2011ApJ...743L...6T,
  2014A&A...567A..34W,2015ApJ...798...96R,2017A&A...599A.108M}. However,
as we discuss in section \ref{s:disc:ss:alt} there is an inconsistency
in the argument of pericenter inferred from mid-IR and scattered light
images. Scattered light images show that the ring is very narrow
($\Delta r / r \approx 0.1$), that the West side of the dust belt is
closer to us, and have quantified the levels of polarisation and
scattering phase function as a function of azimuth
\citep{2015ApJ...799..182P,2017A&A...599A.108M}.
\citet{2012A&A...546A..38L} suggested that the narrow width could be
caused by a planet just exterior to the ring. Another possible
explanation is that the orbits of the planetesimals are dynamically
cold, causing a depletion of the small grains that are normally seen
exterior to the parent belt \citep{2008A&A...481..713T}. This
explanation is particularly relevant here because it predicts that the
disk should have a very small vertical extent, and the dynamical status
of the disk will be a recurring theme throughout.

Thermal emission from HR~4796A's disk has not been imaged at high
spatial resolution at any wavelength, so we obtained ALMA observations
with the goals of i) imaging the population of larger grains that
dominate the emission at millimeter wavelengths, and ii) detecting or
setting limits on any primordial or secondary CO gas. This paper first
presents the ALMA observations and a basic analysis of the continuum and
spectral information contained therein (section \ref{s:obs}). We then
construct and fit disk models with the aim of constraining the disk
structure (section \ref{s:model}), and finish by discussing these
results and placing them in the context of what is already known about
this system (section \ref{s:disc}).

\section{Observations}\label{s:obs}

HR~4796A was observed over two hours by ALMA in band 7 (880$\mu$m)
during Cycle 3 using 41 antennas, with baselines ranging from 15 to
1124~m (2015.1.00032.S). The correlator had 3 spectral windows centered
at frequencies of 333.76, 335.70 and 347.76~GHz each covering a
bandwidth of 2~GHz; these were set up for continuum observations with a
spectral channel width of 15.625~MHz. The remaining spectral window was
centered near the $^{12}$CO J=3-2 line frequency (345.76~GHz) and
covered a bandwidth of 1.875~GHz with a spectral resolution (twice the
channel size, due to Hanning smoothing) of 976.562~kHz (0.85~km s$^{-1}$
at the rest frequency of the line).

The observations were executed in two subsequent scheduling blocks.  The
sources J1427-4206 and J1107-4449 were used as bandpass and flux
calibrators, respectively, and observed at the beginning of each block.
Observations of the science target HR~4796A (50 minutes total
integration time) were interleaved with observations of phase calibrator
J1321-4342 and check source J1222-4122. Calibration and imaging of the
visibility dataset was carried out using the \textsc{CASA} software
version 4.5.2 through the standard pipeline provided by the ALMA
observatory.

We carried out a first round of continuum imaging and deconvolution
using the \textsc{CLEAN} algorithm and Briggs weighting with a robust
parameter of 0.5.  This yields a synthesized beam of size
0.16$\times$0.18\arcsec, corresponding to 11.6$\times$13.1~au at the
72.8~pc distance of the source from Earth.  Given the relatively high
signal-to-noise ratio (S/N) of the emission (peak S/N of 28), we used
the CLEAN model to carry out one round of phase-only self-calibration. A
second round of continuum imaging shows a significant image quality
improvement, now yielding a peak S/N of 37. The standard deviation
obtained near the disk is $\sigma = 31 \mu$Jy beam$^{-1}$, which should
be uniform across the central 2\arcsec~region where the disk is detected
because the primary beam correction in this region is $<$1\%.

In addition to the continuum, we also analysed the high velocity
resolution spectral window around the $^{12}$CO J=3-2 line frequency.
We first subtracted continuum emission in visibility space using the
\texttt{uvcontsub} task within CASA, then imaged the visibilities with
natural weighting to cover the spectral region $\pm 50$~km s$^{-1}$ of
the star's systemic velocity \citep[$\sim$13.7 km/s in the heliocentric
reference frame,][]{2007A&A...474..653V}. This procedure yielded
datacubes with a synthesized beam size of $0.19 \times 0.22$\arcsec at
the native spectral resolution of the dataset (0.85 km s$^{-1}$). The
standard deviation of the noise in the datacube is 2~mJy~beam$^{-1}$ in
a 0.42~km~s$^{-1}$ channel.

\subsection{Basic continuum analysis}\label{s:obs:ss:img}

\begin{figure}
  \begin{center}
    \hspace{-0.5cm} \includegraphics[width=0.5\textwidth]{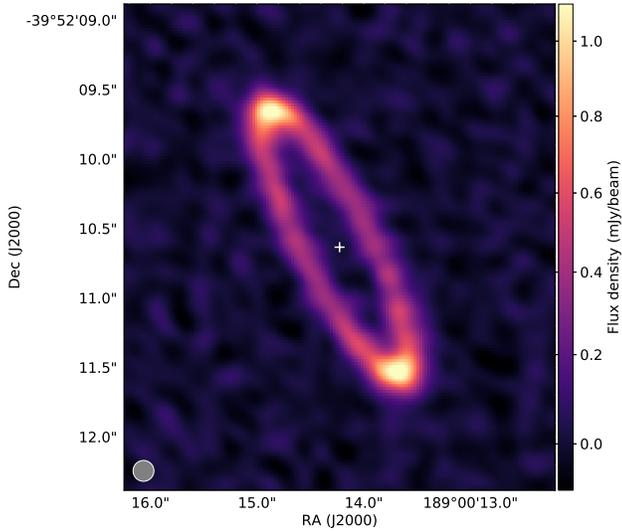}
    \caption{Self-calibrated Briggs-weighted image of the disk around
      HR~4796A (robust = 0.5). The filled circle in the lower left
      corner shows the beam of 0.16$\times$0.18\arcsec. The star is not
      detected but it's location is marked by a +, and with a distance
      of 72.8pc the diameter of the ring is approximately
      160au.}\label{fig:im}
  \end{center}
\end{figure}

We first take a quick look at the observations using the clean image.
Detailed visibility modelling is carried out below, so the purpose of
this section is simply to introduce the data and provide a qualitative
image-based feel for the results that will follow.

Figure \ref{fig:im} shows that the disk is seen very clearly as a narrow
ring that is strongly detected (S/N$>$9) at all azimuths. The width of
the disk appears similar to the beam size of 0.17\arcsec, so given the
distance of 72.8 parsecs the radial and vertical extent of the ring
about the maximum near 80au is no more than about 15au. The star is not
detected, which is consistent with the predicted photospheric flux of
25.5$\mu$Jy.

\begin{figure}
  \begin{center}
    \hspace{-0.5cm} \includegraphics[width=0.5\textwidth]{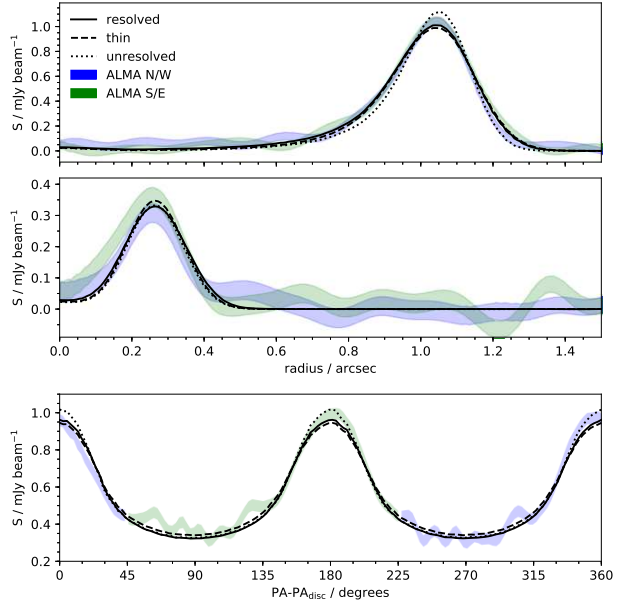}
    \caption{Radial (top two panels) and azimuthal (bottom panel)
      profiles of the surface brightness $S$. The radial profiles are
      along the disk major (upper panel) and minor (middle panel) axes,
      using 10$^\circ$-wide swaths.  The blue transparent bands show
      sections to the N and W, and the green to the S and E; the width
      of these bands is the 1$\sigma$ uncertainty.  The same profiles
      are shown for three different models. The dotted lines show a
      model ring that is `unresolved' radially and vertically.  The
      dashed lines show a vertically `thin' (flat) Gaussian model that
      is resolved radially, and the solid line shows a radially and
      vertically `resolved' model. These models are described in detail
      in section \ref{s:model:ss:gauss}. The unresolved model is a poor
      match to the data, and there is little difference between the
      resolved and thin models.}\label{fig:rad}
  \end{center}
\end{figure}

As a test of whether the ring is resolved, Figure \ref{fig:rad} shows
radial cuts along the major and minor disk axes, and an azimuthal
profile around the disk. For comparison three different models that
provide good fits to the data (and which are described below) are also
shown. Comparing an `unresolved' ring model (dotted line) with the data,
the ring appears clearly resolved in the radial direction, but whether
it is resolved vertically is less clear.  Some clue may be given by the
asymmetry in the inner and outer parts of the radial profile along the
major axis; a vertically `thin' model (dashed line) does not contribute
as much flux as one that is `resolved' both radially and vertically
(solid line), but this difference is barely discernible. Comparison with
the azimuthal profile yields similar results. Thus, while profiles along
both the major and minor disk axes are affected by the radial and
vertical structure, the differences here are small and models of the
full dataset are needed to quantify them.

As a quick test of whether the ring is consistent with being symmetric,
we rotated the image by 180$^\circ$ and subtracted it from the
un-rotated version. The star is not detected, so an x/y shift was
allowed to optimise the subtraction. The result of this subtraction is
an image that appears consistent with noise, suggesting that any
brightness asymmetry that could arise from the disk eccentricity of 0.06
is not detected with ALMA. As the star is undetected, we cannot rule out
the possibility that the disk is eccentric but has an azimuthally
uniform surface density.

Using an elliptical mask with a semi-major axis of 1.75\arcsec\ and
semi-minor axis 0.4\arcsec\ (the ratio derived for the dust ring below),
we measure a total disk flux of $14.8 \pm 1.5$ mJy, where the
uncertainty is dominated by the 10\% absolute calibration
uncertainty. These values are consistent with $14.4 \pm 1.9$mJy measured
with SCUBA-2 \citep{2013MNRAS.430.2513H} as part of the Survey of Nearby
Stars (SONS) legacy programme \citep{2017MNRAS.470.3606H}. This
agreement suggests that the ALMA observations have not resolved out
significant flux on scales larger than seen in Figure \ref{fig:im}.

\subsection{Spectral data and CO}\label{s:obs:ss:co}

The fractional luminosity ($f=L_{\rm disk}/L_\star$) of the disk around
HR~4796A is exceptional (0.5\%), and the system is very young, so we
considered that detection of either remnant primordial or secondary CO
gas in this system was likely with ALMA. However, no clear signal is
detected in the dirty continuum-subtracted data cube. To search more
carefully for secondary CO under the assumption that it is co-located
with the dust, we used the filtering method developed by
\citet{2015MNRAS.447.3936M} as implemented by
\citet{2017ApJ...842....9M}.  In this framework only pixels where the
disk is detected at $>$5$\sigma$ in the continuum are used, and spectra
in each pixel of the imaged data cube are red or blue shifted to account
for the expected radial velocity at that spatial location. This method
assumes the best-fit dust disk geometry derived in section \ref{s:model}
and an estimated stellar mass of 2.18 M$_{\odot}$
\citep{1999A&AS..137..273G}.  This method did not result in a detection
and yields an integrated line flux upper limit (3$\sigma$) of 25 mJy km
s$^{-1}$. A similar search for CO distributed in the same orbital plane
as the disk, but with a different radial extent, also yielded a
non-detection.

\section{Continuum image Models}\label{s:model}

We now place more formal constraints on the disk parameters, modelling
the disk as an optically thin torus using the observed visibilities. To
reduce the computational load we temporally averaged the data into
10-second long chunks, and spectrally averaged the four spectral windows
down to four channels per spectral window. Following averaging, the
visibility weights were recomputed using the CASA \texttt{statwt} task.
This step ensures that the relative visibility weights are correct, but
not necessarily their absolute values,\footnote{For example see
  \href{https://casaguides.nrao.edu/index.php/DataWeightsAndCombination}{https://casaguides.nrao.edu/index.php/DataWeightsAndCombination}.}
and this is corrected below.

The modelling method is the same as used by
\citet{2016MNRAS.460.2933M,2017MNRAS.465.2595M}. For one specific set of
parameters a disk image is first generated using \texttt{radmc-3d}
\citep{2012ascl.soft02015D}. This image is then Fourier transformed to
the visibility plane, where the image is interpolated at the same $uv$
points as our time averaged continuum observations. The difference
between the model and the data is then quantified by computing the
$\chi^2$ goodness-of-fit metric over all visibility samples. In
computing the $\chi^2$ we applied a constant re-weighting factor of
1/2.5 (i.e. we increase the variance by a factor of 2.5) that ensured
the reduced $\chi^2$ for all visibilities was unity \citep[i.e. the
signal from the disk in an individual visibility sample is assumed to be
negligible, which given $N_{\rm vis} = 3210532$ separate visibilities to
be modelled is reasonable, see also][]{2011A&A...529A.105G}. This 
re-weighting ensures that the parameter uncertainties are realistic.
Experiments where this factor was instead included as a model parameter 
find that it is very well constrained ($<$1\% uncertainty), so we chose to
use a constant value for all models.

To find the best fitting model for a given set of parameters, we use the
ensemble Markov-Chain Monte-Carlo (MCMC) method proposed by
\citet{gw2010}, as implemented by the \texttt{emcee} package
\citep{2013PASP..125..306F}. \texttt{emcee} uses an ensemble of
`walkers' (i.e. a series of parallel chains), which are used to inform
the proposals at each step in the chain, increasing the efficiency of
the sampler and allowing for parallel computation. For most fitting runs
we use 40 walkers and chains with 1000 steps, increasing the number of
steps in a few cases with strongly correlated parameters that take
longer to fill out the parameter space. Each model is initialised near
the optimal solution based on prior testing runs, so we typically only
need to discard the first 100 steps as a `burn in' phase.

We tried two families of models: symmetric and asymmetric. The goal of
the symmetric models was to derive best-fit parameters and test whether
the data show evidence for a specific disk radial profile and/or
vertical distribution, and whether different choices for these influence
other parameters. As was suggested in section \ref{s:obs:ss:img} the
disk appears symmetric, so the purpose of an asymmetric model was to
verify that the disk is indeed consistent with being symmetric, and to
quantify the level of asymmetry that could have been detected.

Parameters that are common to all models are the dust mass
$M_{\rm dust}$, the average disk radius $r_0$, the disk position angle
$\Omega$ (measured East of North), the disk inclination $I$, and the
(small) sky offset of the disk from the expected location $x_0$,
$y_0$. The disk is not significantly offset (0.025\arcsec) considering
the $\sim$0.01\arcsec~pointing accuracy of ALMA and the S/N of our
image, which limits the disk eccentricity to less than about 0.05 for a
pericenter direction along the disk major axis (and less than about 0.2
for pericenter along the minor axis). Otherwise these latter two
parameters are unimportant, so feature no further in our analysis. The
data comprise two subsequent observations that are calibrated
separately, so to allow for any differences we include a factor that is
the fractional difference in calibration in the second observation
relative to the first (i.e. we do not consider that the disk brightness
actually changed over one hour at a location where the orbital period is
about 500 years). These seven common parameters define the disk
geometry, scale, and brightness, while the model specific parameters
described below define the detailed radial and vertical structure.

For these models we assume a size distribution of dust from
$D=10$$\mu$m to 1cm with a power-law slope $n(D) \propto
D^{2-3q}$ with $q=11/6$. To compute the opacity needed by 
\texttt{radmc-3d} we use a mix of astronomical silicate, amorphous
carbon, and water ice such that the 880$\mu$m opacity is 0.17~m$^2$~kg$^{-1}$
(45~au$^2$~$M_\oplus^{-1}$). As our observations are in a single narrow
bandpass this choice is arbitrary and the mass is given largely for
comparative purposes (i.e. it has a considerable systematic uncertainty).

\begin{table}
  \caption{Best-fit parameters for the Gaussian, box, and power-law
    models. We follow previous authors' conventions of $\sim$$26^\circ$
    for the ascending node and $\sim$$77^\circ$ for
    inclination. Strictly, for this node the inclination should be
    $\sim$$103^\circ$ because the West side of the disk is closer to us
    (or the node should be $\sim$$206^\circ$ and the inclination
    retained). The dust mass uncertainty includes the contribution from
    the absolute flux calibration}\label{tab:models}
  \begin{center}
    \begin{tabular}{lrrrrrr}
      \hline
      & \multicolumn{2}{c}{Gaussian} & \multicolumn{2}{c}{Box} & \multicolumn{2}{c}{Power-law}\\
      Parameter & Value & 1$\sigma$ & Value & 1$\sigma$ & Value & 1$\sigma$ \\
      \hline
      FWHM$_r$ (au)   & 10 & 1 & -  & - & 5 & 1 \\
      FWHM$_h$ (au)   & 7  & 1 & -  & - & 7 & 1 \\ 
      $\delta_r$ (au) & -  & - & 14 & 1 & - & - \\
      $\delta_h$ (au) & -  & - & 10 & 1 & - & - \\ 
      \hline
      $M_{\rm dust}$ ($M_\oplus$) & 0.35 & 0.04 & 0.35 & 0.04 & 0.35 & 0.04\\
      $r_0$ (au) & 78.6 & 0.2 & 78.4 & 0.2 & 78.4 & 0.2 \\
      $\Omega$ ($^\circ$) & 26.7 & 0.1 & 26.7 & 0.1 & 26.7 & 0.1 \\
      $I$  ($^\circ$) & 76.6 & 0.2 & 76.6 & 0.2 & 76.6 & 0.2 \\
      \hline
    \end{tabular}  
  \end{center}
\end{table}

\begin{table}
  \caption{Best-fit $\chi^2$ values relative to the Gaussian torus model
    (for which $\chi^2 = 3203624.3$). The number of model parameters, and the
    BIC values (relative to the Gaussian model) are also given.}\label{tab:chi2}
  \begin{center}
    \begin{tabular}{lrrr}
      \hline
	  Model & $\Delta \chi^2$ & $N_{\rm param}$ & $\Delta$BIC \\
	  \hline
		Two power law torus & -5.3 & 10 & 9.6 \\
		Power law torus & -5.1 & 9 & -5.1 \\
		Gaussian torus & 0.0 & 9 & 0.0 \\
		Two Gaussian torus & 0.3 & 10 & 15.3 \\
		Eccentric Gaussian torus & 1.2 & 11 & 31.2 \\
		Box torus & 6.6 & 9 & 6.6 \\
		Gaussian torus (`thin') & 21.8 & 8 & 6.8 \\
		Gaussian torus (`narrow') & 65.6 & 8 & 50.6 \\
		Gaussian torus (`unresolved') & 114.2 & 7 & 84.2 \\      
      \hline
    \end{tabular}  
  \end{center}
\end{table}




\subsection{Gaussian torus}\label{s:model:ss:gauss}

\begin{figure}
  \begin{center}
    \hspace{-0.5cm} \includegraphics[width=0.5\textwidth]{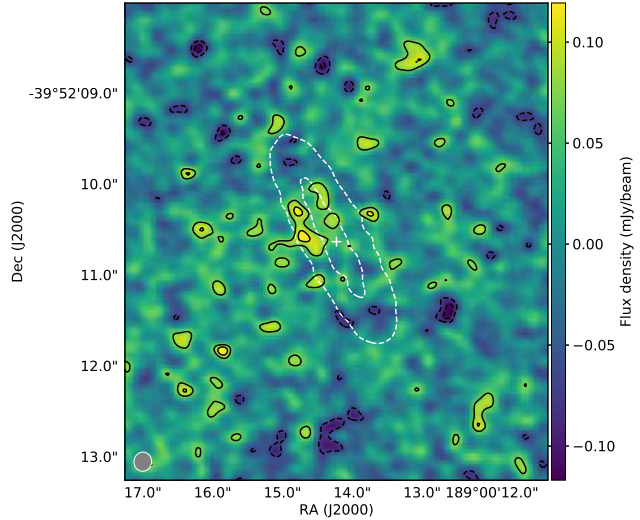}
    \caption{Naturally-weighted image of the residuals after subtracting
      the Gaussian torus model. Solid and dashed contours show the
      residuals at levels of -3, -2, 2, and 3$\sigma$. The star location
      is marked by a +, and white contours show the original image at
      5$\sigma$. A pair of 3$\sigma$ contours remain within the disk
      near the semi-minor axis on the E side.}\label{fig:res}
  \end{center}
\end{figure}

Our `reference' model is a Gaussian torus of radius $r_0$, for which
the additional parameters are the radial $\sigma_r$ and vertical
$\sigma_h$ density dispersions. The full-width of the density at
half-maxima FWHM$_r$ and FWHM$_h$ are therefore
$2\sqrt{2\ln{2}} \approx 2.35$ times larger. The best-fit parameters for
this model are given in Table \ref{tab:models}, and the posterior
distributions for all parameters in Figure \ref{fig:mcmc} in the
Appendix.  A dirty image of the residuals, after subtracting the
best-fit model in visibility space, is shown in Figure
\ref{fig:res}. The overall smoothness of the image shows that the model
is a very good representation of the data. The $\chi^2$ value is
3203624; while this number is not informative in itself, comparison with
the other models, summarised in Table \ref{tab:chi2}, gives it some
context. That is, the difference in $\chi^2$ values between different
models is a more useful indicator of fit quality than the absolute
values, so we quote these relative to this model below
\citep[e.g.,][]{2011A&A...529A.105G}.

Figure \ref{fig:res} shows one interesting feature; a pair of 3$\sigma$
residuals near the disk semi-minor axis on the East side. These appear
for all models, with fluxes of around 100 $\mu$Jy. Inspection of
residual plots for each observation shows that only one blob is present
in each, suggesting that they are either spurious, or fluctuations
caused by noise on top of a larger region of excess flux that is just
below our sensitivity. Their location is near the disk apocenter
inferred from scattered light observations
\citep[e.g.][]{2017A&A...599A.108M}, and whether they provide
constraints on an apocenter glow scenario is considered below.


The basic conclusions from this model are that the disk can indeed be
modelled as a narrow axisymmetric ring, and that the position angle and
inclination are consistent with those derived from scattered light
imaging \citep[][e.g. the latter find $I=76.45 \pm
0.7^\circ$ and $\Omega=27.1 \pm
0.7^\circ$]{2015ApJ...798...96R,2017A&A...599A.108M}. The radii derived
from scattered light appear to differ systematically depending on the
method, for example \citet{2015ApJ...798...96R} find values near 78au
using a 10au-wide elliptical mask, while \citet{2017A&A...599A.108M}
find values near 77au using a locus of the disk's peak brightness. Our
average disk radius is consistent with these results, though agrees more
closely with \citet{2015ApJ...798...96R}, presumably because their
radius estimate is less biased by the
$r^{-2}$ dependence for scattered light.

The radial and vertical extent of the disk is of particular interest
here, and in addition to the search for CO provided the main motivation
for obtaining high resolution images. The fitting results for the
Gaussian model above find that the disk is resolved both radially and
vertically, but given the modest signal to noise ratio seen in Figure
\ref{fig:im}, the lack of significant differences in Figure
\ref{fig:rad}, and that the formal uncertainties on the radial/vertical
extent are about a tenth of the resolution, we made some further tests.

First, we find that the disk is resolved in at least one of the radial
or vertical directions, as an `unresolved' model run where both $FWHM_r$
and $FWHM_h$ were fixed to $<$4au shows significant residuals, primarily
near the ansae ($\Delta \chi^2 = 114.2$). In addition, a `narrow' model
where only $FWHM_r$ is $<$4au (and $FWHM_h$ is allowed to vary) also
shows significant residuals ($\Delta \chi^2 = 65.6$). A `thin' model
where only $FWHM_h$ is $<$4au (and $FWHM_r$ is allowed to vary, yielding
$FWHM_r = 11$au) does not show any significant residuals aside from the
same pair of blobs, but has $\Delta \chi^2 = 21.8$. While a smooth
residual image might result because the disk is not vertically resolved,
it could also arise because the preference for vertical extent is driven
by a low-level signal spread across many beams (as is expected given
that the disk itself spans many beams). The $\Delta \chi^2$ value for
the thin model is higher than for all vertically resolved models
(including the additional models described below), and is more similar
to the value for the narrow case, so a vertically resolved disk is
preferred.

A possibility that we have not yet explored is that a flat disk with a
different radial profile parameterisation could account for the radial
and apparent vertical disk extent. However, a model that has
independently varying inner and outer Gaussian $\sigma_r$
(i.e. $\sigma_{\rm r,in}$ and $\sigma_{\rm r,out}$) still finds a
non-zero $\sigma_h$ (and has $\Delta \chi^2 = 0.3$). We also tested the
possibility that the residual blobs influence the results; adding a
point source to the original Gaussian torus at the location seen in the
first half of the visibility data finds that the disk is still
vertically resolved.


\subsection{Box torus}\label{s:model:ss:box}

As a test of whether a torus with a different structure is also
consistent with the ALMA data, we use a model with uniform space density
within certain radial and vertical limits. A cross section through this
torus yields a rectangular density distribution (i.e. a box), with
radial width $\delta_r$ and vertical height $\delta_h$. While there is
no more motivation for the radial structure than there was for the
Gaussian model, a confined vertical structure could arise if the disk
particles were being perturbed on secular (long) timescales by a
slightly misaligned planet; the total height of the box would be twice
the initial misalignment between the disk and the planet.\footnote{In
  reality the density would actually be higher at the top and bottom of
  the box because the vertical oscillations of an inclined particle are
  sinusoidal. It is for the same reason that the Solar System's
  Asteroidal dust bands are seen as peaks on either side of the ecliptic
  \citep{1984Sci...224...14N}.}

The results for the box model ($\Delta \chi^2 = 6.6$) are similar to the
Gaussian torus (see Table \ref{tab:models}), and again find that the
disk is vertically resolved. Aside from the same peaks to the E of the
star, the residuals are again consistent with noise. Bearing in mind
that the horizontal and vertical extent reported for the box model is
absolute, rather than representative in the Gaussian case, we consider
the results of the two models essentially equivalent (though note that
the $\Delta \chi^2$ is slightly higher for the box model). Thus, while
we can measure the 3-dimensional structure of the disk in terms of the
width and height for both models, we cannot easily discern among
different possibilities for the details of how this dust is distributed
within the torus.

\subsection{Power-law torus}\label{s:model:ss:power}

A final symmetric model retains the Gaussian vertical structure, but has
a radial surface density profile described by a power-law. Specifically,
the density is proportional to
$\left[ (r/r_0)^{-2p_{\rm in}} + (r/r_0)^{2p_{\rm out}} \right]^{-1/2}$.
This profile is regularly used to model scattered light observations,
and more specifically has been applied to the disk around HR~4796A
\citep{1999A&A...348..557A,2017A&A...599A.108M}. By fitting power-laws
to the radial profiles along the disk semi-major axis, the latter
authors found $p_{\rm in} = 23$, and $p_{\rm out} = 13$ to $18$, so one
aim with this model is to test whether the ALMA observations could be
consistent with these parameters. Given the lower spatial resolution of
our ALMA data relative to SPHERE, and the fact that the previous two
models are both adequate descriptions of said ALMA data, we first set
$p = p_{\rm in} = p_{\rm out}$. All other parameters are the same as in
the Gaussian and box tori models.

The best fit power-law index for this model ($\Delta \chi^2 = -5.1$) is
$p = 24 \pm 2$, which corresponds to a FWHM of only 5au. The residuals
are indistinguishable from the results of the previous two models, the
$\Delta \chi^2$ value is slightly lower than the Gaussian torus model,
and again the disk is found to be vertically extended with
$FWHM_h = 7$au.  Relaxing the model to allow $p_{\rm in}$, and
$p_{\rm out}$ to vary independently does not change this conclusion
($\Delta \chi^2 = -5.3$).  While these models are markedly narrower than
the previous ones in terms of FWHM, this narrowness is not actually
detectable with our ALMA resolution and the model width must be driven
by the extended wings in the radial profile. We have nevertheless shown
that the radial profile can be modelled with a power-law profile that is
consistent with the higher spatial resolution scattered light data.

\subsection{Eccentric Gaussian Torus}\label{s:model:ss:ecc}

It is now well established from scattered light imaging that the disk
around HR~4796A has an eccentricity of about 0.06, which at this level
is well approximated as a circular disk whose center is offset from the
star. While the exact magnitude of this offset shows small differences
depending on the dataset and the method used to extract it, the results
are largely consistent (see however section \ref{s:disc:ss:alt} for
further discussion). These observations conclude that the apocenter of
the disk is near the semi-minor axis of the disk on the East side,
slightly below the location of the residual clumps seen in Figure
\ref{fig:res}. To test whether these clumps are indicative of an
apocenter glow model, or constrain the eccentricity that could have been
detected with ALMA, we use the simplified model of
\citet{2016ApJ...832...81P} to prescribe the dust density around an
elliptical annulus. Two additional parameters are required; the
eccentricity of the belt $e$, and the argument of pericenter $\omega$
($\omega$ is measured from the ascending node $\Omega$, so $\omega=0$
corresponds to a pericenter at the NE ansa). Despite the clumps this
model finds that the eccentricity is consistent with zero, with an upper
limit of 0.1 and no preference for any particular pericenter direction,
and therefore shows no evidence for the offset ($\Delta \chi^2 =
1.2$). A probable reason that the apocenter glow model is not favoured
is that the surface brightness should change smoothly around the ring,
while the clumps are relatively localised.

Should we have detected apocenter glow? \citet{2016ApJ...832...81P} note
that the ratio of the disk surface brightnesses at apocenter and
pericenter tends to approximately $1+e$ at long wavelengths where flux
density is linearly dependent on temperature. Thus, based on the
eccentricity derived from scattered light, at most the ratio for
HR~4796A's disk should be about 1.06. The peak S/N in the clean image is
37 per beam (at the ansae), and by experimenting with regions of
different sizes, a peak S/N of 73 was obtained for square regions
0.2\arcsec$^2$ centered on the ansae. A flux difference of
$1/(73/\sqrt{2})=2$\% between the ansae would be detected at 1$\sigma$,
and the sensitivity for other opposing parts of the ring lower because
the fraction of the ring within a given sky area is smaller. Thus,
because the maximum ratio is not necessarily reached at 880$\mu$m, our
non-detection of apocenter glow does not constrain the model.

\subsection{Summary of modelling}

In summary, we find that the dust ring around HR~4796A is strongly
detected with ALMA, and that the parameters of our models are generally
well constrained. All models find the same residual blobs near the
semi-minor axis on the E side of the ring, but we do not consider them
significant and note that their origin might be made clearer with lower
resolution and/or deeper imaging. The ring is clearly radially resolved,
and models where the disk is also vertically resolved yield the lowest
$\chi^2$ values. This conclusion was robust to different models that
might have accounted for an apparent vertical extent with a different
radial profile.  These tests were however not exhaustive.

Formally, we can use the Schwarz criterion (Bayesian Information
Criterion, or BIC) to test which among our models should be preferred
\citep{1978AnSta...6..461S}. This criterion tests whether the
differences in $\chi^2$ values are large enough to be considered
significant, including a penalty for models that have greater numbers of
parameters: $BIC = \chi^2 + N_{\rm param} \ln(N_{\rm vis})$. Differences
in BIC values greater than six should be considered `strong' evidence in
favour of the model with the lower value
\citep{doi:10.1080/01621459.1995.10476572}. The relative BIC values are
given in Table \ref{tab:chi2}, and show that the Gaussian and power-law
torus models are preferred, with preference for the power-law model.
The box torus and vertically unresolved (`thin') models are poor enough
to have `strong' evidence against them, which is despite the thin model
having one less parameter. The BIC imposes a heavy penalty for
additional model parameters because we have a very large number of
visibilities, meaning that the addition of independently varying inner
and outer power law and Gaussian profiles is not well justified given
the small improvement in the fit.  These formal tests largely confirm
what we concluded above. While it remains possible that the disk is not
vertically resolved, the evidence from our modelling suggests that it
is.

Finally, it may be that the clumps are in fact astrophysical, and a sign
that our models are too simple and do not account for underlying
structure that is only marginally detected.  In such a case our
conclusions about the vertical extent could be incorrect because we have
not considered all possible disk models.  In section \ref{s:disc:ss:alt}
we provide some evidence that alternative models merit consideration,
and expect this issue to be resolved with higher resolution imaging.

\section{Discussion}\label{s:disc}

The primary conclusion from our ALMA data is that we have resolved the
debris ring around HR~4796A radially, and probably vertically. In
addition to the requirement of observing at high spatial resolution,
this measurement is made possible by the intermediate inclination of the
disk; we effectively measure the height near the semi-minor axis, and
the width near the ansa, although they can only truly be backed out and
the degeneracy quantified by self-consistent modelling
\citep[see][]{2016MNRAS.460.2933M}. Expressed as full-width half-maxima
from the Gaussian torus model these radial and vertical extents are
respectively 10 and 7au, and 14 and 10au for the box model. Compared to
the disk mean radius of 79au, the radial width can be considered as an
aspect ratio $w={\rm FWHM}_r/r_0=0.13$ (or $\delta_r/r_0=0.16$) and the
vertical extent as a scale height $h={\rm FWHM}_h/(2r_0)=0.04$ (or
$\delta_h/(2r_0)=0.07$). For the box model the height is equivalent to a
maximum particle inclination of 3.5$^\circ$ or opening angle of
7$^\circ$, if particles' ascending nodes are distributed randomly. Using
a power-law radial profile model, we conclude that the width of the disk
as seen with ALMA is consistent with the width seen in scattered light.

The vertical extent of the disk is important for the following
discussion because this extent gives a direct measure of the range of
orbital inclinations of the particles observed. Because the grains
observed by ALMA are large enough to be weakly affected by radiation
pressure, the structure is therefore also representative of larger
bodies.  With the assumption that their nodes are randomly oriented,
these inclinations then set the minimum relative particle velocities and
therefore the level of dynamical excitation in the disk. While the
velocities may be higher if there are also relative radial velocities,
these cannot be inferred from current observations because a ring of
particles on concentric orbits with a range of semi-major axes looks the
same as a ring of particles with a single semi-major axis and a range of
eccentricities and pericenter directions.

This point provides a theoretical reason to be cautious about our
conclusion regarding the vertical extent of the disk. As noted at the
outset, \citet{2008A&A...481..713T} propose that the narrow appearance
of HR~4796A's dust ring in scattered light arises because it is
dynamically very cold (i.e. eccentricities and inclinations less than
0.01). In this case the dust size distribution is depleted in the
smallest ($\sim$10$\mu$m) grains, because their velocities and
destruction rate are increased relative to larger grains by radiation
forces. These small grains typically have eccentric orbits and appear
beyond the parent belt as a `halo', so a disk that lacks them will
appear unusually narrow in scattered light. Such a disk must be
vertically thin, so would not appear to be vertically resolved by our
observations.

While such a scenario may be attractive, and we consider its
implications below, \citet{2008A&A...481..713T} note that a serious
issue is whether such low eccentricties and inclinations can actually be
obtained. The debris disk paradigm requires a reservoir of parent
planetesimals, which inevitably stir the disk to eccentricities and
inclinations of order 0.01 unless they are smaller than a few kilometers
in size.

\subsection{Collisional status}\label{s:disc:ss:coll}

Given that the stellar and disk properties are well known or can be
estimated, the rate at which mass is being lost from the disk can be
calculated with the assumptions that the emitting surface area of the
disk is dominated by the smallest grains, and that these grains are
always destroyed when they collide with each other \citep[Eq. B6
in][]{2017ApJ...842....9M}. The latter assumption requires sufficient
relative velocities between dust grains, which can be obtained in
several ways. If the particle eccentricities are similar to the
eccentricity of the ring and have a range of pericenter directions
(i.e. their orbits are not concentric) the grain-grain collisions are
probably destructive. The same applies if the disk has the vertical
extent suggested by our modelling.  In a very low-excitation scenario
the assumptions become questionable because the smallest dust does not
dominate the dust emission, and lower mass loss rates are possible.

The estimated mass loss rate is 26~$M_\oplus$ Myr$^{-1}$. This rate is
very high compared to estimates for other stars (e.g. using the same
calculation, 0.01 and 0.4 $M_\oplus$ Myr$^{-1}$ for Fomalhaut and
HD~181327 respectively), primarily because HR~4796A's disk has a very
high fractional luminosity $f$ and the mass-loss rate is proportional to
$f^2$. Given that the system age is approaching 10Myr, a prodigious mass
in solids may therefore have been lost since dispersal of the gas disk,
especially if the disk fractional luminosity was higher in the past.
Given the caveat about the excitation level however, this rate could
also be considered as an upper limit.


Comparison of this rate with estimates for the total mass in solids
present is very uncertain, simply because this estimate requires an
extrapolation up to the unknown maximum planetesimal size $D_{\rm c}$
(in km). Using equation (15) from \citet{2008ARA&A..46..339W}, which
assumes a size distribution with $n(D) \propto D^{2-3q}$ and $q=11/6$
(with $M_{\rm tot}$ in units of $M_\oplus$),
\begin{equation}\label{eq:mtot1}
  M_{\rm tot} = f r_0^2 \sqrt{D_{\rm c} D_{\rm bl}} / 0.37
\end{equation}
and again assuming a $D_{\rm bl}=10$ (in $\mu$m) minimum size, yields
$M_{\rm tot} = 270 \sqrt{D_{\rm c}} M_\oplus$ and therefore a mass of
270$M_\oplus$ for a size distribution up to 1km bodies.  Rearranging
equation (16) from \citet{2008ARA&A..46..339W}, which connects the total
mass, maximum planetesimal size, and collisional timescale, yields (with
$t_{\rm coll}$ in Myr):
\begin{equation}\label{eq:mtot2}
  M_{\rm tot} = 140 D_{\rm c} / t_{\rm coll}
\end{equation}
where we have also assumed planetesimal strength $Q_{\rm D}^\star=150$ J
kg$^{-1}$ and eccentricity $e=0.05$ (this model makes various
simplifying assumptions, e.g. that planetesimal strength is independent
of size and that all material resides in a belt of fixed width). In this
model $e$ simply sets the collision velocities, so is interchangeable with 
inclination, and decreasing either results in less frequent collisions and 
a longer collisional lifetime. Thus, if the dynamical excitation is lower, 
so is the inferred disk mass.


Equating (\ref{eq:mtot1}) and (\ref{eq:mtot2}) to eliminate
$M_{\rm tot}$ and solve for $D_{\rm c}$ gives
$D_{\rm c} = 3.6 t_{\rm coll}^2$, from which it can be concluded that
bodies much larger than 1km must be present if the disk has been
grinding down for $t_{\rm coll}$ equal to the system age, otherwise it
would be fainter than observed. If the disk has been evolving for 10Myr
up to 360km bodies are needed, corresponding to a disk mass of
5000$M_\oplus$. For a collisional evolution time of only 1Myr, 4km
bodies could be colliding, and the total disk mass 500$M_\oplus$. If we
assume $e=0.01$, 2km bodies are needed and the disk mass is
350$M_\oplus$.

To put these estimates in perspective, a 26$M_\oplus$ `isolation mass'
object \citep{1987Icar...69..249L} would form from material within a
similar radial extent as the ring around HR~4796A, and corresponds to a
surface density of 0.1 g cm$^{-2}$ ($2.7 M_\oplus$ au$^{-2}$), similar
to the solid component of the `minimum mass Solar nebula' at this
distance \citep{1977Ap&SS..51..153W}. Similar surface densities have
also been estimated for protoplanetary disks
\citep[e.g.][]{2009ApJ...700.1502A}.

These very high disk masses may present a problem; if the disk is not
dynamically cold the collision rates are such that requiring a
reasonable disk mass ($<$100$M_\oplus$, say, remembering that all of
this mass is confined to the observed ring) requires that the largest
planetesimals be smaller than kilometers in size, but the lifetime of
the disk at the observed level is then much shorter than the system age
($<$1Myr). Conversely, requiring that the disk be able to survive at the
observed level for a sizeable fraction of the system age requires large
($>$100km) planetesimals, and therefore a very large disk mass. This
mass problem is not unique to HR~4796A, and possible solutions arise
when assumptions made above are relaxed, such as the strength of the
planetesimals and their size distribution, that dust only originates in
collisions, or that the systems have been colliding for shorter than the
apparent stellar age. See \citet{2017arXiv171103490K} for a general
discussion of this issue.


As noted above, both the disk mass and mass loss rate problems are
related to the vertical extent of the disk, and both are also alleviated
if the disk is dynamically very cold. In addition, the smaller
planetesimals required would stir the disk less, meaning that the low
excitation could be consistent with the expected level of stirring from
the planetesimals \citep[though whether a lack of $\gtrsim$1km
planetesimals is consistent with planet formation models is debatable,
see][]{2017arXiv171103490K}. The mass and mass loss rate issues might
therefore be resolved if higher resolution observations showed that the
disk is in fact thinner than our modelling suggests.

\subsection{CO gas}\label{s:disc:ss:comass}

\subsubsection{CO mass upper limit}

In section \ref{s:obs:ss:co}, we derived an upper limit to the observed
integrated line flux of the $^{12}$CO J=3-2 transition for gas
co-located with the debris ring. We now translate this flux into an
upper limit on the total CO mass in the belt and aim to understand the
origin of any CO that may still be present below our sensitivity limit.

To quantify the implications of this limit, we calculate the population
of the upper level of the transition (J=3) with respect to all other
energy levels of the CO molecule, using an improved version of the
non-local thermodynamic equilibrium (NLTE) analysis of
\citet{2015MNRAS.447.3936M} that now includes the effect of fluorescence
excitation (Matr\`a et al. submitted).

To calculate collisional excitation, we assume the main collider species
to be electrons, as they have been shown to be the most likely to
dominate collisions with CO in second-generation gas
\citep[e.g.][]{2016MNRAS.461..845K, 2017MNRAS.464.1415M}; collision
rates are obtained from \citet{1975JPhB....8.2846D}. Regardless, the CO
mass derived from our flux upper limit is independent of our choice of
collisional partner \citep[e.g.][]{2015MNRAS.447.3936M}.

To calculate radiative excitation, we consider the radiation field
impinging on a CO molecule at the debris ring's center, including
stellar emission at UV to IR wavelengths (affecting electronic and
vibrational transitions), as well as dust continuum and CMB emission at
far-IR to mm wavelengths (affecting rotational transitions). The stellar
emission is taken as that of a 9650K PHOENIX model atmosphere
\citep[][the temperature derived by fitting to optical
photometry]{2005ESASP.576..565B}, whereas the dust continuum radiation
field is measured assuming our best-fit dust model at 0.88 mm and
scaling it to other far-IR/mm wavelengths using the observed SED. The
PHOENIX models are of the stellar photosphere, so the UV emission could
be higher. However, HR~4796A was detected between 1500 and 3000nm by the
UV Sky-Survey Telescope in the TD-1A satellite
\citep{1973MNRAS.163..291B}.\footnote{VizieR catalogue II/59B}.  Aside
from one value that is about 20 percent higher, the fluxes are
consistent with our photosphere model, so there is no evidence of a
significant UV excess.

We then proceed to solve the system of equations of statistical
equilibrium to obtain the fractional population of our level of interest
($x_{\rm J=3}$) as a function of the unknown electron density (which we
varied between 10$^{-3}$ and 10$^{12}$ cm$^{-3}$) and kinetic
temperature (which we varied between 10 and 250 K). Finally, we assume
that CO emission is optically thin and use Eq. 2 from
\citet{2015MNRAS.447.3936M} to derive a CO mass upper limit from our
observed integrated flux upper limit, again as a function of electron
density and kinetic temperature. We find a CO mass upper limit ranging
between (1.2 to 3.7) $\times$ 10$^{-6}$ M$_{\oplus}$, where this range
is effectively independent of the electron density assumption, and only
weakly dependent on our already wide range of temperatures assumed; we
therefore adopt 3.7 $\times$ 10$^{-6}$ M$_{\oplus}$ as the strict upper
limit on the CO mass derived from our data.

\subsubsection{Primordial origin of undetected CO excluded}

To assess whether any undetected CO could be left over from the
protoplanetary phase of evolution, we need to consider whether i) such a
low CO mass could be optically thick to the line of sight, causing our
CO mass upper limit to be underestimated and ii) whether CO could have
survived photodissociation from the central star since the end of the
protoplanetary phase of evolution. In order to do so we draw an analogy
with the Fomalhaut ring, since it has a similar radial extent and
inclination to that of HR~4796A \citep{2017ApJ...842....8M}, leading to
a similar column length that a CO molecule in the center of the ring
`sees' towards the star ($\sim$6.5~AU), and a similar column length of
CO throughout the ring along the line of sight to Earth ($\sim$13~AU).
Assuming a uniform density torus, the maximum CO number density in the
HR~4796A ring is 2.1 cm$^{-3}$, leading to maximum column densities of
2.0 and 4.0$\times$10$^{14}$ cm$^{-2}$, respectively.

Using Eq. 3 from \citet{2017MNRAS.464.1415M}, for the whole range of
electron densities and kinetic temperature considered above, we set an
upper limit to the optical thickness of the $^{12}$CO J=3-2 line along
the line of sight to Earth of $\tau_{\rm 345 GHz}\leq 0.4$. This shows
that our optically thin assumption is a good approximation and most
likely valid for any CO co-located with the debris ring. Furthermore,
the results of \citet{2009A&A...503..323V} indicate that the maximum
column density of CO along the line of sight to the star leads to very
little self-shielding against photodissociating UV photons; even when
including the shielding effect of a potential primordial H$_2$ reservoir
with a low CO/H$_2$ ratio of 10$^{-6}$, the increase in CO lifetime
against photodissociation is only a factor $\sim$5.

Using the same model stellar spectrum as above, and the modified
\citet{1978ApJS...36..595D} interstellar UV field of
\citet{2008arXiv0806.0088V}, together with photodissociation
cross-sections from \citet{2009A&A...503..323V}, we derive a
photodissociation timescale of eight years at the radial location of the
ring's centre. HR~4796A is an A0-type star, and as such has a relatively
high UV luminosity, so the lifetime of CO is much shorter than the 120
years typically assumed when CO dissociation is driven solely by
interstellar ultraviolet photons
\citep{2009A&A...503..323V,2015MNRAS.447.3936M,2017MNRAS.469..521K}.  We
therefore conclude that any CO present in the HR~4796A ring below our
detection threshold cannot have survived for more than $\sim$40 years,
ruling out the hypothesis that primordial CO gas could have survived
since the protoplanetary phase of evolution.

While these estimates are based on CO that is restricted to be
co-located with the dust (expected because the CO lifetime is much
shorter than the orbital period at that distance), we estimate that the
lifetime of a broader distribution would still be very short. If we
assume a CO disk with the same number density as our upper limit that
extends all the way to the star, the radial column density would be a
factor $\approx$10 higher.  For the same CO/H$_2$ ratio assumed above
the photodissociation time therefore increases by a factor of a few, but
is still significantly shorter than the age of the system.

\subsubsection{The CO+CO$_2$ ice reservoir in HR~4796A's exocomets}

Given the short lifetime of any CO that is co-located with the debris
ring, any such gas that exists below our detection limit must originate
in the planetesimals that feed the observed dust. Other studies have
used the steady-state mass-loss rate from the collisional cascade, in
concert with a CO detection or upper limit and a CO lifetime, to
estimate or set limits on the fraction of CO and CO$_2$ ice in the
parent planetesimals.

Taking our measured CO mass upper limit, the derived CO lifetime of
eight years indicates a CO mass loss rate of $<0.46$ M$_{\oplus}$
Myr$^{-1}$. In steady state, this rate can be combined with the
estimated mass loss rate from the collisional cascade (section
\ref{s:disc:ss:coll}) to measure an upper limit of $<$1.8\% on the
CO+CO$_2$ ice mass fraction in exocomets within the HR~4796A ring. This
fraction is lower than the CO+CO$_2$ mass fractions estimated for Solar
System comets and the Fomalhaut system
\citep[see][]{2017ApJ...842....9M}, but comparable with the estimate for
the debris ring around the F2 type star HD~181327
\citep{2016MNRAS.460.2933M}. As noted above, this fraction is uncertain
because it relies on the uncertain dust mass loss rate, and could
therefore be higher if the disk is vertically thin (in which case this
mass loss rate is considered an upper limit, and could be much
lower). As before, quantifying the vertical extent of the disk can
resolve this issue.

\subsection{Radial width and comparison with scattered light}\label{s:disc:ss:sc}

\begin{figure}
  \begin{center}
    \hspace{-0.5cm} \includegraphics[width=0.5\textwidth]{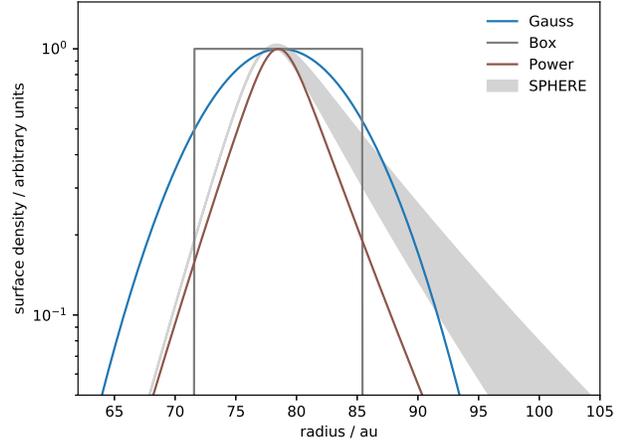}
    \caption{Radial profiles of the main models from section
      \ref{s:model} (solid lines) and the 1/$r^2$-corrected scattered
      light flux profile from SPHERE observations (grey line and filled
      region). The spatial resolution of our ALMA observations is about
      the same as the width of the box model (FWHM = 12.4au). The
      vertical range of the plot is chosen to approximately reflect the
      signal to noise ratio of the observations.}\label{fig:prof}
  \end{center}
\end{figure}

One of the primary drivers for obtaining these data was to compare the
radial distributions of larger grains, as seen with ALMA, with the
smallest grains, as seen in scattered light. For this comparison, we use
the results of \citet{2017A&A...599A.108M}, who measure a FWHM$_r$ of
7au using a power-law model (by measuring the width along the semi-major
axis). As shown in section \ref{s:model:ss:power} the best fitting
power-law model is consistent with that derived from the scattered light
data. However, as highlighted by the range of radial widths derived from
the models, our resolution is insufficient to say whether the dust as
seen by ALMA is as radially concentrated as it appears in scattered
light. The diversity of possibilities is illustrated in Figure
\ref{fig:prof}, which shows the radial profiles of the Gaussian, box,
and power-law models. For comparison, the power-law fit to the SPHERE
data along the disk's major axis is also shown, where the range covered
by the gray filled region shows the difference between the outer
profiles seen towards the NE and SW ansae \citep{2017A&A...599A.108M}.

We conclude that the radial extent of the smallest grains in the disk as
imaged by SPHERE appears to be very similar to that for larger
$\sim$mm-sized grains as imaged by ALMA. This similarity is unexpected,
because dust near the blowout limit should reside on high-eccentricity
orbits, creating a `halo' beyond their source region that has a
scattered light surface brightness power-law profile of $r^{-3.5}$
\citep{2006ApJ...648..652S,2006A&A...455..509K,2008A&A...481..713T}.  As
noted above, one explanation could be the low planetesimal excitation
scenario, while a related possibility is that the ring is radially
optically thick. Another scenario is shepherding by an outer planet
\citep{2012A&A...546A..38L}.

The outer shepherding planets considered by \citet{2012A&A...546A..38L}
had masses in the range 3 to 8$M_{\rm Jupiter}$, which, aside from
uncertainties in the conversion between mass and brightness, are not
favoured by more recent direct imaging \citet{2017A&A...599A.108M}, so
we consider this possibility unlikely.

Considering the high radial optical depth scenario (which prevents small
dust from leaving the ring before it is destroyed),
\citet{2008A&A...481..713T} find that radial optical depths of order
unity are required for the halo to be significantly attenuated. If the
disk is vertically resolved then the radial optical depth is
$f r_0 / h \approx 0.1$, and it seems implausible that the radial
optical depth in the HR~4796A disk is sufficiently high to be
responsible for the lack of a small-grain halo. Similarly, if our
measurement of the vertical scale height for the disk is correct, low
planetesimal excitation is implausible and would rule out this
possibility. Thus, in addition to having implications for the uncertain
disk mass and mass loss rate, further mm-wave observations can help
understand the role of radial optical depth and dynamical excitation in
setting the steep radial profile seen in scattered light.

\subsection{Expectations from secular perturbations}\label{s:disc:ss:sp}

What do the various measurements mean, if anything, for the history and
status of the debris ring? If we assume that the offset seen in
scattered light and the pericenter glow seen in the mid-infrared arise
from a planet-induced (`forced') eccentricity $e_{\rm f}$ within the
disk of about 0.06 (but see section \ref{s:disc:ss:alt} for discussion
of this assumption), then the present appearance of the ring depends on
the initial conditions, which we now discuss. We then consider how
secular perturbations set constraints on the putative planet's mass and
semi-major axis. If the disk is vertically resolved further constraints
are possible, because the lifetime of the disk as it currently appears
is inferred to be short.

\subsubsection{Initial conditions}

Considering the vertical extent first, any initial misalignment between
the planet and the disk causes the bodies' ascending nodes to precess.
The precession rate is a function of semi-major axis, so for a disk of
finite width differential precession eventually randomises the nodes of
neighbouring planetesimals, and the final vertical extent of the disk is
twice the initial misalignment.  Thus, our inferred vertical extent
could arise from a very flat disk initially inclined 3.5$^\circ$
relative to the planet. However, the vertical extent could equally arise
because this was the intrinsic range of inclinations, but this scenario
requires that any initial planet-disk misalignment was very small. The
way to distinguish between these possibilities is to measure the
vertical density distribution; in the former scenario the density will
be highest at the highest inclinations \citep[][see also section
\ref{s:model:ss:box}]{1984Sci...224...14N}, while for the latter the
density is almost certainly more concentrated towards the mid-plane. If
the disk is vertically very thin, and a planet causes the disk
eccentricity, then any initial misalignment was very small.

The argument for the radial extent is similar, but any scenario must
also satisfy the observed eccentricity. Secular perturbations from a
planet with semi-major axis $a_{\rm pl}$ impose a forced eccentricity
and longitude of pericenter\footnote{Note that longitude of pericenter,
  which is the argument of pericenter plus the longitude of the
  ascending node $\omega+\Omega$ is appropriate here because the bodies'
  nodes may be regressing (i.e. precession due to misalignment with the
  planet's orbit). See \citet{1999ApJ...527..918W} for a detailed
  description of the dynamics of pericenter glow.} $\varpi_f$; the orbit
of any body that already has these values for $e$ and $\varpi$ will not
change, while the $e$ and $\varpi$ of any other body will change such
that the highest eccentricity occurs when the pericenter is aligned with
the planet's (along $\varpi_f$). Thus, the width of a disk can either
reflect the disk's initial width, as long as the initial eccentricity
happened to be at the forced location, or the disk could have initially
been narrow with low eccentricity and the width mostly contributed by
pericenter precession. By `mostly', we mean that the width cannot be
solely contributed by pericenter precession because if all planetesimals
were all initially at the same semi-major axis $a$, then there would be
no differential precession and the disk would only ever be a narrow ring
whose pericenter precesses. A finite width means that differential
precession due to different semi-major axes can eventually randomise the
orbits (`phase mix' in $e \cos \varpi$, $e \sin \varpi$ space) and
pericenter glow set up \citep{2005A&A...440..937W}. If we simply assume
that the initial width is narrower than the observed width
($\lesssim$5au), then the width expected from precession is about
$2ae_f$, which is similar to that measured.

Which of these origins is more likely? The fact that the disk width is
close to that expected given an initial distribution that was both
narrow and on circular orbits may favour this initial
condition. However, it also seems possible that the orbits of a
population of planetesimals orbiting exterior to a planet could `relax'
to the forced values due to some dissipative process, the prime
candidate being gas drag before and during gas disk dispersal. In either
case, the presence of an exterior planetesimal population might be the
result of a `pile up' of dust in the gas pressure maximum just
external to a planet \citep{2012A&A...545A..81P}, and the most likely
initial conditions predicted by further development of such
models. Occam's razor suggests that the planet that caused the pile-up,
and the planet causing the observed disk to be offset from the star, are
one and the same.

\begin{figure}
  \begin{center}
    \hspace{-0.5cm} \includegraphics[width=0.5\textwidth]{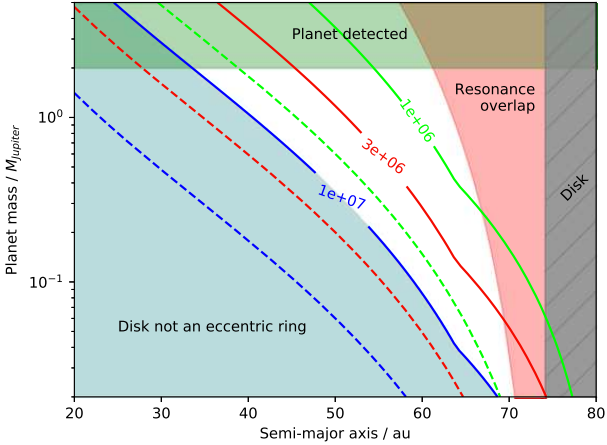}
    \caption{Limits on locations of an interior planet that imposes the
      eccentricity of the HR~4796A debris ring. The solid contours show
      the mass and semi-major axis of planets that cause the disk to
      appear eccentric after the times given by each label. The dashed
      lines show the mass and semi-major axes for the onset of
      collisions at the same times. Planets above about 2
      $M_{\rm Jupiter}$ would have been detected, planets too close to
      the disk would eject particles via resonance overlap, and planets
      too far away would not force the ring to be eccentric within the
      lifetime of the star. The equations used to generate this plot are
      given in Appendix \ref{app:mpl}}\label{fig:mpl}
  \end{center}
\end{figure}

\subsubsection{Planet constraints}

What kind of planet could impose the structure on the disk? Continuing
with the picture of an interior planet, the primary requirements are i)
that the planet's semi-major axis and eccentricity result in a forced
eccentricity $e_{\rm f} = 0.06$ at 79au (see Appendix \ref{app:mpl}),
and ii) that the planetesimals have undergone sufficient precession at
79au within some time (e.g. the age of the system, or the time elapsed
since gas-disk dispersal). Based on a rough maximum initial width of 5au
(see above), we quantify `sufficient' by requiring that planetesimals
2.5au exterior to 79au have precessed through at least one full cycle,
and that planetesimals 2.5 interior to 79au are at least one precession
cycle ahead of those at the outer edge. Given the discussion above about
the relation between initial and final disk widths, this condition is an
approximation, but does not significantly affect our conclusions. This
differential precession condition is very similar to the orbit-crossing
criterion of \citet{2009MNRAS.399.1403M}, the main difference being that
particles need not precess a full cycle farther than their neighbours
for their orbits to cross.

In general, the closer the planet to the disk, the more rapidly the disk
is affected, so which condition dominates the phase-mixing requirement
depends on the planet location; differential precession is slower than
outer-edge precession when the planet is more distant from the
disk. However, a planet cannot lie arbitrarily close to the disk, as it
would remove bodies on short timescales, and therefore must lie farther
than required by the resonance-overlap criterion
\citep{1980AJ.....85.1122W}.

The resonance-overlap and precession criteria, plus an approximate
planet detection limit of 2 Jupiter-masses \citep{2017A&A...599A.108M},
are shown in Figure \ref{fig:mpl} (the equations used to generate this
plot are given in Appendix \ref{app:mpl}). Shaded regions at the upper
and right boundaries of the figure show the regions of parameter space
that are ruled out by resonance-overlap and planet detection limits.
The solid lines show contours along which sufficient precession has
occurred within 1, 3, and 10Myr. For times greater than 10 Myr, the disk
has not precessed enough to appear as a uniform eccentric ring, which
provides the third diagonal criterion to the lower left that bounds the
planet location.

Where might an interior planet reside? The longer the disk has been in a
state where secular perturbations can act as assumed (i.e. since gas
disk dispersal) the lower the mass and the farther from the disk this
planet can be. Given the estimated system age near 10Myr, the
appropriate contour in Figure \ref{fig:mpl} in this scenario probably
lies between 3 and 10Myr, meaning that roughly speaking the putative
planet should be more massive than Neptune, and within 40au of the disk.

So far the constraints on this supposed planet have been purely set by
dynamics, considering the time taken for the debris ring to appear as it
does given plausible initial conditions. If the disk is vertically thin,
it could be dynamically cold and collisions may be relatively
unimportant (i.e. there are no mass loss rate or disk mass problems),
and no more constraints are possible. However, if the disk has the
vertical extent suggested by our modelling then the disk lifetime at the
current brightness should be very short, and the inferred disk mass very
large. This problem can be alleviated if the onset of collisions was
relatively recent, which is possible if these collisions were initiated
by the same perturbations that cause the disk to appear eccentric.

To this end, Figure \ref{fig:mpl} also shows contours of constant
collision-onset times \citep[using the method outlined by][but without
the assumption of $a_{\rm pl}/r_0 \ll 1$]{2009MNRAS.399.1403M}. These
assume that collisions begin when disk particles have precessed
sufficiently that their orbits overlap. For a given planet collisions
begin well before the disk has precessed to the point that it appears
smooth and eccentric, so the dashed contours are well below the solid
ones. The onset of collisions is sufficiently short that the disk would
have been losing mass for essentially the entire time taken for secular
perturbations to make the disk appear eccentric. Thus, to avoid the disk
mass problem the disk should have acquired the eccentric structure
recently, so that the time since the onset of collisions is also short.
This requirement does not necessarily mean that the gas-rich phase of
disk evolution only ended recently, as the planet may have obtained an
orbit necessary to stir and perturb the disk some time well after gas
dispersal (e.g. by interaction with a second planet). Regardless, the
preferred current planetary parameters are in the upper right of the
allowed space; within 10-20au of the disk inner edge and with a mass
similar to Jupiter.


This disk lifetime problem may also be alleviated by allowing the orbit
of the planet to acquire the eccentricity necessary to stir and perturb
the disk some time well after gas dispersal, for which a probable
mechanism would be interaction with a second planet. Such a scenario is
less attractive because of the added complexity, but is hard to rule
out.

\subsection{Alternative scenarios}\label{s:disc:ss:alt}

While the eccentric nature of the ring has led to secular perturbation
induced pericenter-glow being the favoured interpretation for the ring
around HR~4796A, there remain issues with this interpretation that we
now outline. The need for reconciliation of these issues points to
alternative hypotheses.

The pericenter-glow hypothesis is based on mid-IR observations
\citep{1999ApJ...527..918W,2000ApJ...530..329T,2011A&A...526A..34M},
which have relatively low spatial resolution. The relevant
observable is therefore the flux ratio between the two ansae. The
constraints on the disk's forced eccentricity and pericenter are
degenerate, with the derived eccentricity being minimal (about 0.06)
when the pericenter is at the NE ansa \citep{2011A&A...526A..34M}. The
eccentricity can be higher, but to ensure the brightness asymmetry does
not become too great the pericenter must be moved away from the
ansa. This degeneracy is not total however, and
\citet{2011A&A...526A..34M} show that it can be broken by considering
the temperature profile along the disk semi-major axis.  Using this
constraint they conclude that the pericenter is near the NE ansa, though
quote an uncertainty of 30$^\circ$. However, for pericenters that are
far from the ansa the eccentricity becomes much larger than 0.06, for
example 0.3 when $\omega=\pm 75^\circ$.\footnote{These models, which do
  not consider the effect of the disk eccentricity on collision rates as
  a function of azimuth, probably underestimate the disk eccentricity
  required; \citet{2017arXiv170408085L} find that including these
  effects decreases the strength of pericenter glow.}


The disk offset can be measured directly through high resolution
imaging. In this case an ellipse with a non-zero stellocentric offset is
fitted to the disk image, and the resulting parameters de-projected to
yield the orbital elements of the ring. Several measurements have been
made for this offset using scattered light imaging, which consistently
find a small but non-zero eccentricity \citep[$<$0.1,
][]{2009AJ....137...53S,2011ApJ...743L...6T,
  2014A&A...567A..34W,2015ApJ...798...96R,2017A&A...599A.108M}. The
arguments of pericenter vary somewhat, but are consistently closer to
the semi-minor axes than they are to the ansa (suggesting that the
scattering phase function is not influencing the results), and not
consistent with a pericenter near the NW ansa. These results are
therefore in conflict with those derived from pericenter glow in the
mid-IR.

While these details do not point to specific alternative scenarios, they
force us to consider relaxing modelling aspects that are commonly
implicit in most debris disk models. One possible resolution is that the
point in the disk that is closest to the star is indeed near the
semi-minor axis on the Western side of the star, but the dust actually
tends to be brighter near the NW ansa. To ensure that the $\sim$10K
colour difference between the NE and SW ansae seen in the mid-IR is also
satisfied, merely increasing the amount of dust would not be
sufficient. The simplest explanation is that the dust at the NE ansa
tends to be smaller, and therefore hotter and brighter than elsewhere in
the disk. Such a difference in grain sizes might also be associated with
an increase in space density, as an increased collision rate at a more
dense disk location could cause an increase in the amount of small dust
near that location.

Thus, an alternative picture for the disk around HR~4796A essentially
involves decoupling the disk brightness and the geometry. The enhanced
brightness and dust temperature at the NE ansa results from dust that is
smaller on average. Secular perturbations may still be invoked for the
ring eccentricity, but other ideas are possible. For example, the bulk
of the dust in the ring may be the result of a single previous
collision, which has since spread into a largely, but not entirely,
uniform ring \citep{2014MNRAS.440.3757J}. In this scenario, the ring
retains the orbit of the progenitor, providing the origin of the ring's
eccentricity, and collisions are more frequent at the original collision
location, which explains the increased dust temperature near the NE
ansa.

The development of such alternative models is not the goal of this
paper, but comparison of the mid-IR and scattered light results, and
considering the implications of the collisional status of the system,
suggests that there is sufficient evidence that their exploration is
well motivated. The residuals seen in Figure \ref{fig:res}, near the
apocenter as derived from scattered light (the semi-minor axis on the
Eastern side), may provide further motivation, though further
observations to test whether they are real are desirable.

\section{Summary and Conclusions}\label{s:concl}

We have presented the first high spatial resolution mm-wave images of
the debris disk around the young star HR~4796A, revealing a narrow ring
of roughly mm-sized grains. Modelling of the radial and vertical
structure with a variety of axisymmetric models shows that we have
resolved the disk radially, with a $\sim$10au extent that is consistent
with that seen in scattered light. These models consistently find that
the disk is also vertically resolved, with a similar extent. Residual
images show that these models provide a very good fit to the data; the
only remaining structure is a few 3$\sigma$ blobs near the semi-minor
axis on the East side of the star. We remain cautious about the claim of
vertically resolved structure because it is smaller than the beam size,
but find that it is robust to models that use a range of different
radial profiles.

Various solutions have been proposed for the narrowness of the disk in
scattered light. One that seems promising is a low excitation scenario
that preferentially depletes the smallest dust
\citep{2008A&A...481..713T}. This scenario is attractive because it
provides a way to lower estimates of high disk mass and mass loss
rates. However, this scenario is in conflict with our conclusion that
the disk is vertically resolved, so higher resolution observations that
confirm or refute the vertical extent would be very valuable for
considering the dynamical excitation and collisional status of the belt.

We do not detect any CO gas, and rule out the possibility that any
remaining undetected CO gas is primordial. Under the assumption that the
disk is vertically resolved, we set an approximate limit on the
CO+CO$_2$ ice fraction in the parent planetesimals of $<$1.8\%. This
value is at the low end of abundances observed in Solar System comets
and other similarly aged exocometary belts, but could be higher if the
disk has low dynamical excitation because the the mass loss rate used in
the estimate would be lower.

We consider a scenario where the disk eccentricity arises from secular
perturbations from an interior planet. Such a planet may be the reason
the ring exists, having trapped radially drifting dust just exterior to
its orbit during the gas-rich phase of evolution. Using constraints that
bound its location and mass, we find that such a planet should be more
massive than Neptune, and lie exterior to 40au.

Finally, we highlight a conflict between the interpretations of mid-IR
and scattered light observations. While both suggest ring eccentricities
of about 0.06, the former argues for a pericenter near the NE ansa,
while the latter consistently finds the pericenter near the semi-minor
axis on the West side of the star. These conclusions do not appear
reconcilable, so we suggest that models that allow the spatial dust
density and grain size to vary as a function of azimuth, independently
of the pericenter location, should be considered.

\section*{Acknowledgments}

We thank the referee for a careful reading of the manuscript. 
GMK is supported by the Royal Society as a Royal Society University
Research Fellow. LM acknowledges support from the Smithsonian
Institution as a Submillimeter Array (SMA) Fellow. OP is supported by
the Royal Society as a Royal Society Dorothy Hodgkin Fellow.

This paper makes use of the following ALMA data:
ADS/JAO.ALMA\#2015.1.00032.S. ALMA is a partnership of ESO (representing
its member states), NSF (USA) and NINS (Japan), together with NRC
(Canada), MOST and ASIAA (Taiwan), and KASI (Republic of Korea), in
cooperation with the Republic of Chile. The Joint ALMA Observatory is
operated by ESO, AUI/NRAO and NAOJ.

This publication uses (among others) the python packages APLpy
\citep{2012ascl.soft08017R}, astropy \citep{2013A&A...558A..33A}, corner
\citep{2016JOSS.2016...24F}, matplotlib \citep{2007CSE.....9...90H},
numpy, and scipy \citep{2011arXiv1102.1523V}.

\appendix

\section{Modelling results}

\begin{figure*}
  \begin{center}
    \hspace{-0.5cm} \includegraphics[width=1\textwidth]{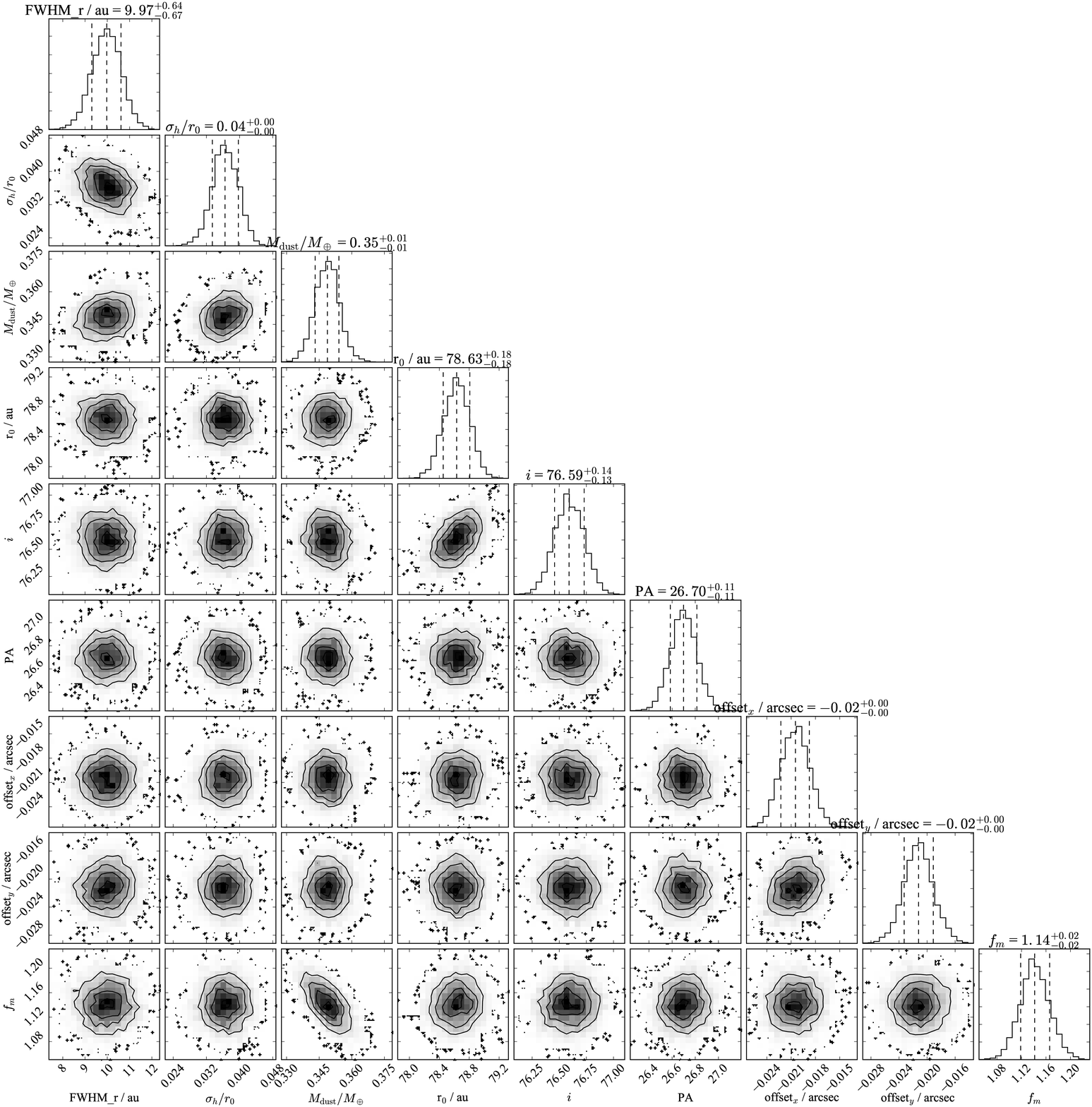}
    \caption{An example showing the posterior distributions of
      parameters from the MCMC fitting, in this case for the Gaussian
      torus model. In off-diagonal panels the solid lines show 1, 2, and
      3$\sigma$ contours and the greyscale shows the density, and in the
      diagonal panels the histograms show the 1D distribution for each
      parameter, titled by the median and $\pm$1$\sigma$
      uncertainty.}\label{fig:mcmc}
  \end{center}
\end{figure*}

Figure \ref{fig:mcmc} shows posterior distributions for all parameters
from the `reference' Gaussian torus model of section
\ref{s:model:ss:gauss}.  The distributions show that the parameters are
well constrained and show little degeneracy.


\section{Planet constraints}\label{app:mpl}

This section details the constraints used to generate Figure
\ref{fig:mpl}. Values assumed are $r_0 = 79$au, $e_f = 0.06$, and
$M_\star = 2.18 M_\odot$

\subsection{Resonance overlap}

The region marked `Resonance overlap' is set by the resonance
overlap criterion of \citep{1980AJ.....85.1122W}, at which point:
\begin{equation}
  a_{\rm pl} = r_{\rm in} \left[ 1 - 
    1.3 \left(\frac{M_{\rm pl}}{M_\star}\right)^{2/7} \right]
\end{equation}
where the inner disk edge $r_{\rm in}$ is assumed to be $r_0$ minus half
the observed Gaussian disk width of 10au.

\subsection{Secular perturbations}

The eccentricity $e_{\rm pl}$ of a planet with semi-major axis $a_{\rm pl}$
that results in planetesimals at $r_0$ with eccentricities $e_f$ is given by
\begin{equation}
  e_{\rm pl} = e_f b_{3/2}^{(1)}(\alpha) / b_{3/2}^{(2)}(\alpha)
\end{equation}
where $b_s^{(j)}$ are Laplace coefficients and $\alpha = a_{\rm pl}/r_0$.

These Laplace coefficients can be written as:
\begin{equation}
  b_{3/2}^{(1)} = 3 \alpha \mathcal{F}(3/2,5/2,2,\alpha^2)
\end{equation}
\begin{equation}
  b_{3/2}^{(2)} = 15 \alpha^2 \mathcal{F}(3/2,7/2,3,\alpha^2) / 4
\end{equation}
where $\mathcal{F}$ is the standard hypergeometric function.

The secular precession frequency of planetesimals at $r$ under the
influence of this planet is
\begin{equation}
  A(r) = \frac{n}{4} \frac{M_{\rm pl}}{M_\star} 
    \frac{a_{\rm pl}}{r} b_{3/2}^{(1)}(\alpha)
\end{equation}
where $n = \sqrt{G M_\star / r^3}$ is the mean motion at $r$.

The precession time at the outer edge is then $2 \pi / A(r_{\rm out})$,
and the differential precession time between the inner and outer disk
edges $2 \pi / [A(r_{\rm in}) - A(r_{\rm out})]$. Because the precession
widens the disk, we use $r_{\rm in} = r_0 - 2.5$au and
$r_{\rm out} = r_0 + 2.5$au here (i.e. an approximate initial width,
which is narrower than the observed width). The lines in Figure
\ref{fig:mpl} show the greater of these two quantities.

The crossing timescale given by \citet{2009MNRAS.399.1403M} is used, but
without the simplifying assumption that $a_{\rm pl} \ll r$:
\begin{equation}
  t_{\rm cross} = -1 / (e_f \, r \, dA/dr)
\end{equation}
where we use a numerical derivative for $dA/dr$.


\end{document}